\documentclass[12pt]{article}
\usepackage{epsfig}
\newcommand\slv{v\kern-5pt\raise1pt\hbox{$\scriptstyle/$}\kern1pt}

\def\be{\begin{equation}}
\def\ee{\end{equation}}
\def\bq{\begin{eqnarray}}
\def\eq{\end{eqnarray}}
\def\l{\langle}
\def\r{\rangle}
\newcounter{saveeqn}
\newcounter{App} %\setcounter{App}{0}

\begin{document}
\thispagestyle{empty}
\begin{flushright}
NIKHEF-00-020\\
UM-TH-00-18\\
%(Version of~\today)\\
\end{flushright}
\vspace{0.5cm}

\begin{center} 
{\Large \bf New approach for calculating heavy-to-light form }\\[3mm]
{\Large \bf factors with QCD sum rules on the light-cone}\\[.3cm]
\vspace{1.7cm}
{\sc \bf Stefan Weinzierl$^{1,a}$ and Oleg Yakovlev$^{2,b}$}\\[1cm]
\begin{center} \em 
$^1$ NIKHEF, P.O. Box 41882, NL - 1009 DB Amsterdam, The Netherlands\\
$^2$ Randall Laboratory of Physics, University of Michigan, Ann Arbor\\
   Michigan 48109-1120, USA
\end{center}\end{center}
\vspace{2cm}

% abstract ---------------------------------------
\begin{abstract}\noindent
{
We suggest a new approach for calculating heavy-to-light form factors. 
The method is based on light cone sum rules (LCSR) and covers
the whole kinematical range of momentum transfer.
The derivation of the new sum rule uses a suitable
combination of double and single dispersion
integrals.
As an example we give numerical results for the form factor $f^+$ for the 
$B \rightarrow \pi$ transition.
}    
\end{abstract}
\vspace*{\fill}

% footnotes -------------------------------------
\noindent $^a${\small e-mail: stefanw@nikhef.nl}\\
\noindent $^b${\small e-mail: yakovlev@umich.edu}

% main text ------------------------------------
\newpage

\section{Introduction}

The accurate study of B meson decays is a main source of information 
for understanding $CP$ violation. 
We expect that the upcoming experiments will measure a variety of 
B decay properties with good precision \cite{BaBar}.  
In order to over-constrain the unitary triangle one pursues not only a measurement
of the various angles, but tries as well to determine the length of the sides of the
triangle.
One important quantity is the CKM matrix element $|V_{ub}|$, which is proportional to the
length of one side of the unitary triangle.
$|V_{ub}|$ can be obtained from semileptonic B decays.
There are two complementary strategies for a determination of $|V_{ub}|$, relying either on 
exclusive or inclusive measurements. 
The inclusive decay $B\to X l\nu$ can be calculated using the heavy-quark 
expansion \cite{Inclusive,Uraltsev}. 
However one of the main obstacles here 
is that experimentally it is mandatory to impose restrictive cuts to suppress
the background from $ B\to X_cl\nu$ decays. 

On the other hand an exclusive measurement through the channels $B \rightarrow \pi l \nu$
or $B \rightarrow \rho l \nu$ is easier from an experimental point of view.
It requires however a reliable theoretical calculation of the heavy-to-light
transition form factors.
These form factors parameterize the relevant hadronic 
matrix elements and are non-perturbative quantities.
They can be calculated using lattice methods \cite{lattrev}-\cite{JLQCD} 
or QCD sum rules on the light-cone.  
In this paper we focus on the sum rule approach.

The QCD light-cone sum rule method (LCSR)  
has been suggested in \cite{BraunLC,CZhLC,BF} 
and is a combination of the operator product expansion (OPE) on
the light-cone \cite{Lepage,Efremov,Chernyak} with QCD sum rule
techniques \cite{SVZ}.  For a review of the method and 
results we refer the reader to \cite{Reviews,Reviews2}.

Within the sum rule approach results for the heavy-to-light form 
factors corresponding to the $B \rightarrow \pi$, 
$B \rightarrow K$, $D \rightarrow \pi$ or $D \rightarrow K$
transitions are easily obtained from one master formula \cite{BKR}-\cite{Ball1}. 
Changing the light meson from a pseudo-scalar meson 
to a vector meson (e.g. $\rho$, $\omega$, $K^\ast$ or $\phi$) the form 
factors of semi-leptonic or rare radiative decays of a $B$-meson into 
vector mesons are obtained \cite{Ball1,vec1,vec2}. 

To illustrate our method we focus on the
form factors relevant to the $B \rightarrow \pi$ transition,
defined through
\bq
\label{definitionF}
\l \pi(q) | \bar{u} \gamma_\mu b | B(p+q) \r & = & 2 f^+(p^2) q_\mu 
  + \left( f^+(p^2) + f^-(p^2) \right) p_\mu,
\eq
where $p+q$, $q$ and $p$ denote the $B$ and $\pi$ four-momenta and the momentum transfer,
respectively, and $f^\pm$ are the two independent form factors.
If one neglects lepton masses only the form factor $f^+$ is relevant.
The standard sum rule technique starts from the correlation function of two
heavy-light currents
\bq
\label{correlationfunction}
F_\mu(p,q) & = & i\int dx e^{ip\cdot x}<\pi(q)|T\{\bar u(x)\gamma_\mu b(x),
m_b\bar b(0)i\gamma_5 d(0)\}|0>.
\eq
The correlation function is then expanded near the light-cone $x^2 \approx 0$.
Each power of $x^2$ in the light-cone expansion leads to an additional power of
$m_b^2-(p+uq)^2$ in the denominator. 
Here $u$ denotes the momentum fraction carried by one of the quarks inside the pion.
The light-cone expansion is justified,
provided that $(p+q)^2$ and $p^2$ are sufficiently smaller than $m_b^2$.
In particular this implies that
$p^2 < m_b^2 - 2 m_b \chi$, where $\chi$ is a scale
of order $\Lambda_{QCD}$.
This limits the application of the sum rule to the kinematic
region to 
\bq\label{regLCSR}
0 < p^2 < m_b^2 - 2 m_b \chi.
\eq
However, in order to extract $|V_{ub}|$ one needs a prediction for the
form factor $f^+$ in the whole kinematical range $0 < p^2 < (m_B -
m_\pi)^2$, which is much wider than (\ref{regLCSR}). 
Several attempts have been made in the past to overcome this problem.
One of the first solutions \cite{Bagan} consisted of extrapolating
the form factor from the region $0 < p^2 < m_b^2 - 2 m_b \chi$ to
the region $m_b^2 - 2 m_b \chi < p^2 < (m_B - m_\pi)^2$ by assuming the functional form
\bq
\label{fit}
f^+(p^2) & = & \frac{f^+(0)}{1-a p^2/m_{B^\ast}^2+ b (p^2/m_{B^\ast}^2 )^2}.
\eq
Another analysis used the fact that
close to the point $p^2=m_{B^\ast}^2$ the form factor can be described 
by a simple pole model (see for example \cite{Voloshin,Burdman,Grinstein}),
assuming vector meson dominance:
\bq
\label{polemodel}
f^+(p^2) & = & \frac{f_{B^\ast} g_{B^\ast B \pi}}{2 m_{B^\ast} \left( 1 - p^2/m_{B^\ast}^2 \right)}
\eq
The coupling $g_{B^\ast B \pi}$ can be calculated by QCD sum rules techniques, starting from
the same correlation function eq.~(\ref{correlationfunction}) and using a double dispersion relation.
This additional information was used in \cite{Reviews2} by fitting a function of the form of eq.~(\ref{fit})
to the low $p^2$-behavior obtained from the standard light-cone sum rule and to a single
pole model description for high $p^2$.
A third approach was used in the analysis of \cite{K2000}: The Becirevic-Kaidalov parameterization
\cite{Becirevic}
\bq
\label{BK}
f^+(p^2) & = & c_B \left( \frac{1}{1-p^2/m_{B^\ast}^2} - \frac{\alpha}{1-p^2/(\gamma m_{B^\ast}^2)} \right)
\eq
was assumed as functional form and the parameters of this ansatz were determined from QCD sum rules,
using the light-cone sum rule for $0 < p^2 < m_b^2 - 2 m_b \chi$ as well as the information on the
coupling $g_{B^\ast B \pi}$.
In the parameterization eq.~(\ref{BK}) the first term corresponds to the vector meson pole, whereas the second
term represents an effective contribution from all higher resonances.

All of the methods mentioned above involve ad hoc assumptions, which are difficult to justify from first principles.
In addition we cannot easily quantify the error associated with these additional  
assumptions. 

In this letter we suggest a new method for calculating heavy-to-light
form factors which 
yields a prediction in the whole kinematical range of momentum transfer and overcomes the problem outlined above.
We start from the same correlation function eq.~(\ref{correlationfunction}).
The derivation of the new sum rule is based on a combination of double and single dispersion
integrals.
The particular combination of double and single dispersion integrals ensures that our sum rule
is valid over the whole kinematical region.
As input data we need the value of $f^+(0)$, together with the first $l$ derivatives $f^{+(l)}(p^2)$
at $p^2=0$ (with $l$ an integer), which can be obtained (numerically) from the standard sum rule for $f^+(p^2)$.
In addition we need the value of the coupling $g_{B^\ast B \pi}$, which can be obtained from the sum rule
for the coupling.
We derive the new sum rule to the same accuracy to which 
the two other sum rules are known \cite{Bpi,Coupling,Bagan}:
to next-to-leading order in twist 2 and to leading order in twist 3 and 4.
We present numerical results for the form factor $f^+(p^2)$ for the 
$B \rightarrow \pi$ transition.

This paper is organized as follows:
In the section 2 we introduce the new sum rule.
In section 3 we give for the new sum rule 
the QCD corrections to twist 2 relevant for the $B \rightarrow \pi$ transition.
Numerical results are given in section 4.
Section 5 contains our conclusions.

\section{The formalism}

We write the correlation function eq.~(\ref{correlationfunction}) in terms of invariant amplitudes
$F(p^2,(p+q)^2)$ and $\tilde{F}(p^2,(p+q)^2)$
\bq
F_\mu(p,q) & = & 
F(p^2,(p+q)^2) q_\mu + \tilde{F}(p^2,(p+q)^2) p_\mu,
\eq
and focus on $F(p^2,(p+q)^2)$.
We denote by
\bq
\sigma(p^2,s_2) & = & \frac{1}{\pi} \; \mbox{Im}_{s_2} \; F(p^2,s_2), \nonumber \\
\rho(s_1,s_2) & = & \frac{1}{\pi^2} \; \mbox{Im}_{s_1} \; \mbox{Im}_{s_2} \; F(s_1,s_2)
\eq
the imaginary part of $F$ with respect to $s_2$ and the double imaginary part
of $F$, respectively.
Furthermore we define by
\bq
{\cal B}_{p^2} & = & \lim\limits_{p^2 \rightarrow -\infty, n \rightarrow \infty, -p^2/n=M^2}
\frac{1}{(n-1)!} \left( - p^2 \right)^n \left( \frac{d}{dp^2} \right)^n
\eq
the Borel operator with respect to the variable $p^2$.
Hadronic representations for the spectral densities $\sigma(p^2,s_2)$ and $\rho(s_1,s_2)$
are written by singling out the ground state:
\bq
\sigma^{hadr}(p^2,s_2) & = & 2 m_B^2 f_B f^+(p^2) \delta(s_2-m_B^2) + \sigma^{hadr}(p^2,s_2) \Theta(s_2-s_0), \nonumber \\
& & \nonumber \\
\rho^{hadr}(s_1,s_2) & = & m_B^2 m_{B^\ast} f_B f_{B^\ast} g_{B^\ast B \pi} \delta(s_1 - m_{B^\ast}^2)
\delta(s_2 - m_B^2)  \nonumber \\
& & + \rho^{hadr}(s_1,s_2) \left(1-\Theta(s_1,s_2)\right) \Theta(s_1-m_b^2) \Theta(s_2-m_b^2).
\eq
Here $\Theta(s_1,s_2)$ defines the duality interval in the $(s_1,s_2)$-plane 
for the ground state. A convenient
choice is given by the square
\bq
\Theta(s_1,s_2) & = & \Theta(s_0-s_1) \Theta(s_0-s_2)
\eq
The standard sum rule for the form factor $f^+(p^2)$ is obtained from writing a single 
dispersion relation for $F(p^2,(p+q)^2)$ in the $(p+q)^2$-channel, inserting the
hadronic representation and Borelizing in $(p+q)^2$:
\bq
\label{eqa1}
{\cal B}_{(p+q)^2} F(p^2,(p+q)^2) & = & {\cal B}_{(p+q)^2} \left( \frac{2 m_B^2 f_B f^+(p^2)}{m_B^2-(p+q)^2} + \int\limits_{s_2>s_0} ds_2 \frac{\sigma^{hadr}(p^2,s_2)}{s_2-(p+q)^2} \right)
\eq
The Borel operator ensures that any subtraction terms which might appear will vanish after
Borelization.
One proceeds to replace $\sigma^{had}$ by $\sigma^{QCD}$ and equate the r.h.s of eq.~(\ref{eqa1})
to the QCD calculation of ${\cal B}_{(p+q)^2} F(p^2,(p+q)^2)$. This yields the standard
sum rule for $f^+(p^2)$:
\bq
\label{eqa1a}
f^+(p^2) & = & \frac{1}{2m_B^2f_B} \int\limits_{m_b^2}^{s_0} \sigma^{QCD}(p^2,s_2)
e^{-\frac{s_2-m_B^2}{M^2}}ds_2
\eq
In a similar way the standard light-cone sum rule for the coupling $g_{B^\ast B \pi}$ is 
obtained from a double dispersion
relation:
\bq
\label{eqa2}
{\cal B}_{p^2} {\cal B}_{(p+q)^2} F(p^2,(p+q)^2) & = &  
{\cal B}_{p^2} {\cal B}_{(p+q)^2} \left( \frac{m_B^2 m_{B^\ast} f_B f_{B^\ast} g_{B^\ast B \pi}}
{(p^2 - m_{B^\ast}^2)( (p+q)^2 - m_B^2)} \right. \nonumber \\
& & \left. + \int\limits_{\Sigma} ds_1 ds_2 \frac{\rho^{hadr}(s_1,s_2)}{(s_1-p^2)(s_2-(p+q)^2)} \right),
\eq
where $\Sigma$ denotes the union of $s_1>s_0$, $s_2>m_b^2$ 
with $s_1>m_b^2$, $s_2>s_0$. 
Again the presence of the Borel operators ensures that any subtraction will give a vanishing 
contribution.
The LCSR for the coupling $g_{B^\ast B \pi}$ reads:
\bq
\label{eqa2a}
g_{B^\ast B \pi} & = & \frac{1}{m_B^2 m_{B^\ast} f_B f_{B^\ast}}
\int\limits_{m_b^2}^{s_0} ds_1 \int\limits_{m_b^2}^{s_0} ds_2 
\rho^{QCD}(s_1,s_2) e^{- \frac{(s_1-m_{B^\ast}^2)+(s_2-m_B^2)}{M^2}}
\eq
Here we took the two Borel parameters to be equal.\\
\\
To derive our new sum rule we suggest here to use a dispersion relation 
for $\sigma(p^2,s_2)/(p^2)^l$ in the $p^2$-channel (with $l$ being an integer)
\bq
\label{eqa3}
\sigma(p^2,s_2) & = & - \frac{1}{(l-1)!} \left(p^2\right)^l \frac{d^{l-1}}{ds_1^{l-1}}
\left. \frac{\sigma(s_1,s_2)}{s_1-p^2} \right|_{s_1=0} 
+ \int\limits_{s_1>m_b^2} ds_1 \frac{(p^2)^l}{s_1^l} \frac{\rho(s_1,s_2)}{s_1 -p^2} 
\eq
and to replace $\sigma(p^2,s_2)$ in eq.~(\ref{eqa1}) by the r.h.s of eq.~(\ref{eqa3}).
By choosing $l$ high enough the dispersion relation eq.~(\ref{eqa3}) will be convergent.
We obtain
\bq
\label{eqa4}
\lefteqn{
{\cal B}_{(p+q)^2} F(p^2,(p+q)^2) = } & & \nonumber \\
& & {\cal B}_{(p+q)^2} \left(
\frac{2 m_B^2 f_B f^+(p^2)}{m_B^2-(p+q)^2} + \int\limits_{s_1>m_b^2, s_2>s_0} ds_1 ds_2
\frac{(p^2)^l}{s_1^l} \frac{\rho(s_1,s_2)}{(s_1-p^2)(s_2-(p+q)^2)} \right. \nonumber \\
& & \left. - \frac{1}{(l-1)!} \int\limits_{s_2>s_0} \frac{ds_2}{s_2-(p+q)^2} (p^2)^l
\left. \frac{d^{l-1}}{ds_1^{l-1}} \frac{\sigma(s_1,s_2)}{s_1-p^2} \right|_{s_1=0}
\right)
\eq
Furthermore we write down a double dispersion relation for $F(p^2,(p+q)^2)/(p^2)^l$:
\bq
\label{eqa5}
\lefteqn{
{\cal B}_{(p+q)^2} F(p^2,(p+q)^2) = } & & \nonumber \\ 
& & {\cal B}_{(p+q)^2} \left( \frac{(p^2)^l}{(m_{B^\ast}^2)^l} 
\frac{m_B^2 m_{B^\ast} f_B f_{B^\ast} g_{B^\ast B \pi}}
{(p^2 - m_{B^\ast}^2)( (p+q)^2 - m_B^2)} 
+ \int\limits_{\Sigma} ds_1 ds_2 \frac{(p^2)^l}{s_1^l} \frac{\rho(s_1,s_2)}{(s_1-p^2)(s_2-(p+q)^2)} \right. \nonumber \\
& & \left. - \frac{1}{(l-1)!} \int\limits_{s_2>m_b^2} \frac{ds_2}{s_2-(p+q)^2} (p^2)^l
\left. \frac{d^{l-1}}{ds_1^{l-1}} \frac{\sigma(s_1,s_2)}{s_1-p^2} \right|_{s_1=0} \right)
\eq
Again, by choosing $l$ high enough the dispersion integral will be convergent in the
$s_1$-channel. The Borel operator ensures that any subtraction terms in the $s_2$-channel
will vanish.
Now equating the r.h.s of eq.~(\ref{eqa4}) with the r.h.s of eq.~(\ref{eqa5}) we obtain
the sum rule
\bq
\label{eqa6}
f^+(p^2) & = & \frac{1}{2} \frac{(p^2)^l}{(m_{B^\ast}^2)^l} \frac{f_{B^\ast} g_{B^\ast B \pi}}
{m_{B^\ast} \left( 1 -\frac{p^2}{m_{B^\ast}^2} \right)} 
- \frac{1}{(l-1)!} \left( p^2 \right)^l \left. \frac{d^{l-1}}{ds_1^{l-1}} 
\frac{f^+(s_1)}{s_1-p^2} \right|_{s_1=0} \nonumber \\
& & + \frac{1}{2 m_B^2 f_B} \int\limits_{\Sigma'} ds_1 ds_2 
\frac{(p^2)^l}{s_1^l} \frac{\rho(s_1,s_2)}{s_1-p^2} e^{- \frac{s_2-m_B^2}{M^2}},
\eq
where the region $\Sigma'$ is defined by $s_1>s_0$ and $m_b^2 < s_2 < s_0$.
This sum rule is valid in the whole kinematical range of $p^2$. 
As input data we need the first $(l-1)$ terms of the
Taylor expansion of $f^+(p^2)$ around $p^2=0$. 
These parameters can be obtained numerically  from
the standard sum rule eq.~(\ref{eqa1a}).
We further need the residuum at the pole $p^2=m_{B^\ast}^2$, which can be obtained from
the sum rule eq.~(\ref{eqa2a}).
The new sum rule agrees by construction with the standard sum rule eq.~(\ref{eqa1a}) in a Taylor
expansion around $p^2=0$ up to the first $(l-1)$ terms. Furthermore the residuum at $p^2=m_{B^\ast}^2$ agrees with
the coupling sum rule eq.~(\ref{eqa2a}).

We remark that the parameter $l$ plays a similar role as the Borel parameter $M^2$:
There is a lower limit on $l$ since the dispersion relations eq.~(\ref{eqa3}) and eq.~(\ref{eqa5})
have to converge.
Going to higher values for $l$ will improve the convergence of the dispersion relations
and suppress higher resonances in the $B^\ast$-channel.
But there is also an upper limit on $l$: The higher the value of $l$, the more derivatives
of $f^+(p^2)$ at $p^2=0$ we have to know. At some point we start probing 
the region $p^2 > m_b^2 - 2 \chi m_b$, at which the standard sum 
rule eq.~(\ref{eqa1a}) might break down.
By using the sum rule with various values of $l$, say $l=1,2,3$ and by looking at the variation of the
results, we can get an estimate of the uncertainty of our method.

In the case $l=0$ the second term is absent. For $l=0$ the first term
corresponds to the pole model. 
As we will show explicitly below, the last term 
vanishes for the leading order twist 2, 3 and 4 contributions.
The first non-vanishing contribution comes from the $\alpha_s$-corrections to the
twist 2 contribution.
This might explain the empirical fact that for some 
heavy-to-light transitions (like for $D\to\pi$) the simple
pole model approximates the form factor reasonably well.

\section{The additional term}

We now consider the last term in eq.~(\ref{eqa6})
\bq
\label{eqa7}
f^+_{corr}(p^2) & = & \frac{1}{2m_B^2f_B}  
\int\limits_{\Sigma'} ds_1 ds_2 \left( \frac{p^2}{s_1} \right)^l 
\frac{\rho(s_1,s_2)}{s_1-p^2}
e^{- \frac{s_2-m_B^2}{M^2}},
\eq
which has to be evaluated. 
The sum rules eq.~(\ref{eqa1a}) and eq.~(\ref{eqa2a}) are 
known in twist 2 to NLO accuracy, and in twist 3 and 4 to LO. 
Aiming at the same accuracy for eq.~(\ref{eqa7}) we find that
the  LO contributions of twist two, three and four vanish.
This is due to the fact that the spectral density $\rho(s_1,s_2)$ at leading order
is localized along the diagonal $s_1=s_2$, whereas the integration region $\Sigma'$ lies
beyond the diagonal.
At next-to-leading order the situation is different. The radiative 
corrections give contributions which are smeared over the whole $s_1-s_2$
plane, overlapping with the region $\Sigma'$.
For the twist 2 NLO contribution we find:
\bq\label{fcorr}
\lefteqn{
f^+_{corr}(p^2) = 3 C_F \frac{\alpha_s}{\pi}
\frac{m_b^2 f_\pi}{2 m_B^2 f_B} \left(\frac{p^2}{m_b^2}\right)^l
\exp\left(-\frac{m_b^2-m_B^2}{M^2}\right) 
\left\{ \left[
\int\limits_1^\infty dr \frac{g_1(r)}{(r-1)^3} 
\right. \right. } & & \nonumber \\ 
& & \left. \left.
- \int\limits_1^\infty dr \frac{g_1'(1)}{(r-1)^2} 
- \frac{1}{2} \int\limits_1^2 dr \frac{g_1''(1)}{r-1} \right] 
+2 \left[ \int\limits_1^\infty dr \frac{g_2(r) \ln(r)}{(r-1)^3}
-\int\limits_1^2 dr \frac{g_2'(1)}{r-1} \right] \right\}
\eq
where
\bq
g_1(r) & = & -\int\limits_{\frac{1+r}{r} z_0}^{(1+r) z_0} ds \; r \;e^{-\frac{bs}{1+r}}
\left( \frac{1+r}{1+r+rs} \right)^l \frac{1}{1+r+rs -a -ar} \nonumber \\
& & \cdot \left( \frac{s}{1+r+s} + \ln\left( \frac{1+r+rs}{1+r+s}\right) \right), \nonumber \\
g_2(r) & = & \int\limits_{\frac{1+r}{r} z_0}^{(1+r) z_0} ds \; r \;e^{-\frac{bs}{1+r}}
\left( \frac{1+r}{1+r+rs} \right)^l \frac{1}{1+r+rs -a -ar}.
\eq
Here we used the dimensionless variables 
\bq
a = \frac{p^2}{m_b^2}, \; b = \frac{m_b^2}{M^2}, \; z_0 = \frac{s_0-m_b^2}{m_b^2}.
\eq
Eq.~(\ref{fcorr}) is written in such a form that singularities at
$r\to 1$, which are present in individual contributions, cancel
explicitly in the combinations inside the square brackets. 
The values of $g_1$ and $g_2$ and their derivatives at $r=1$ are 
\begin{eqnarray}
g_1(1) & = & 0, \qquad
g_1'(1)  =  - e^{-bz_0} \frac{1}{(1+z_0)^l} \frac{z_0^2}{1+z_0} 
\frac{1}{1+z_0-a}, \nonumber \\
g_1''(1) &= & - e^{-bz_0} \frac{1}{(1+z_0)^l} \frac{z_0^2}{(1+z_0)^2} 
\frac{1}{(1+z_0-a)^2}\Big( l z_0 a -2 z_0 a -b z_0 a  \nonumber \\
 & &-b z_0^2 a - a + b z_0 + b z_0^3 + 2 b z_0^2 
  + z_0^2 - l z_0^2 -l z_0 + 1 + 2 z_0 \Big), \nonumber \\
g_2(1) & = & 0, \qquad
g_2'(1)  =  - e^{-bz_0} \frac{1}{(1+z_0)^l} \frac{z_0}{1+z_0-a}. 
\end{eqnarray}
The integrations over $r$ and $s$ can be performed numerically. 

\section{Numerical results}

\begin{figure}
\centerline{
\epsfig{
file=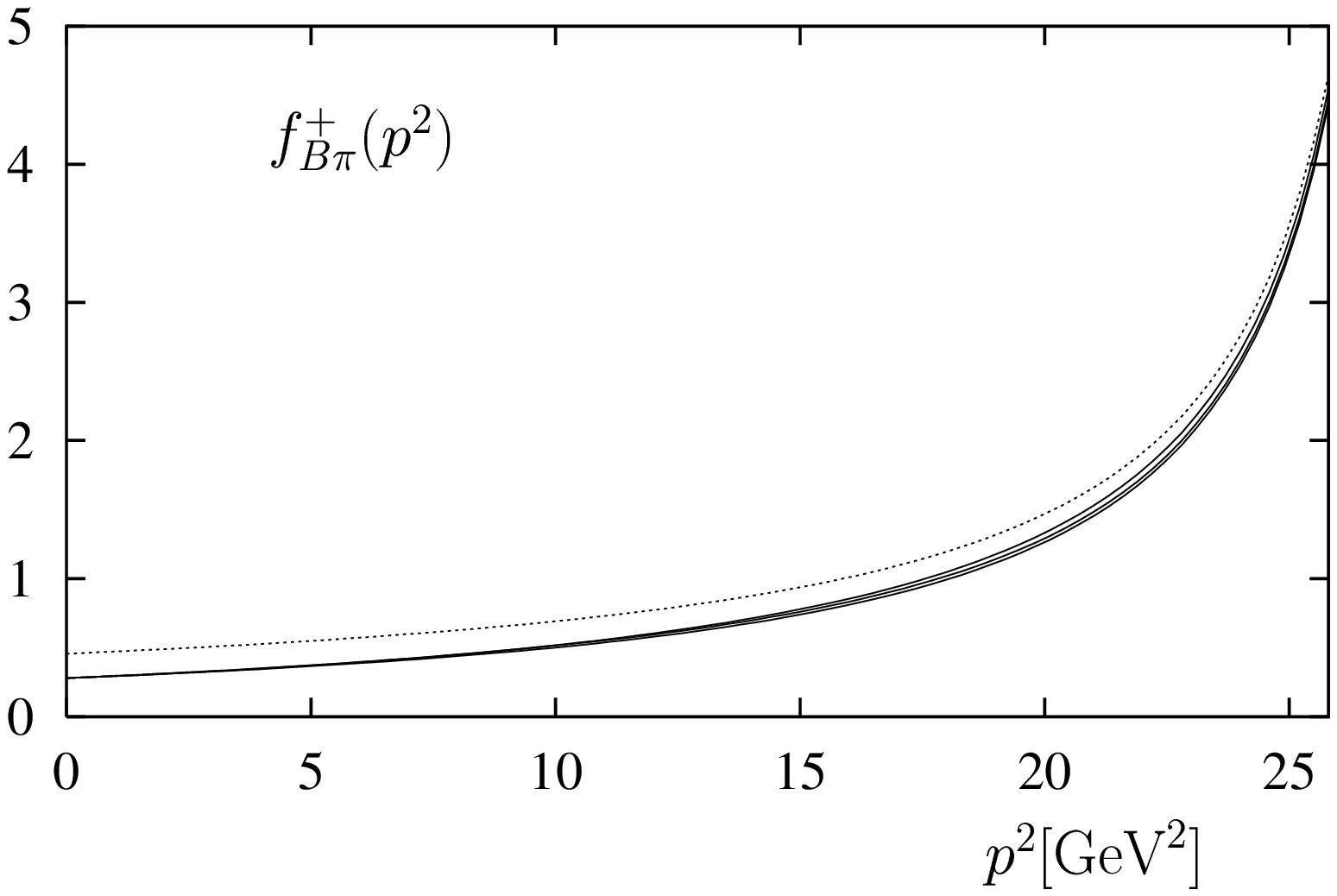, %plot.ps
scale=0.9,%
clip=}}
\caption{\label{fig1} 
The results of the new sum rule for the form factor $f^+_{B\to\pi}(p^2)$ 
for different numbers $l$.
The dotted line corresponds to $l=0$, the solid lines correspond to $l=1,2,3$.}
\end{figure}

\begin{figure}
\centerline{
\epsfig{file=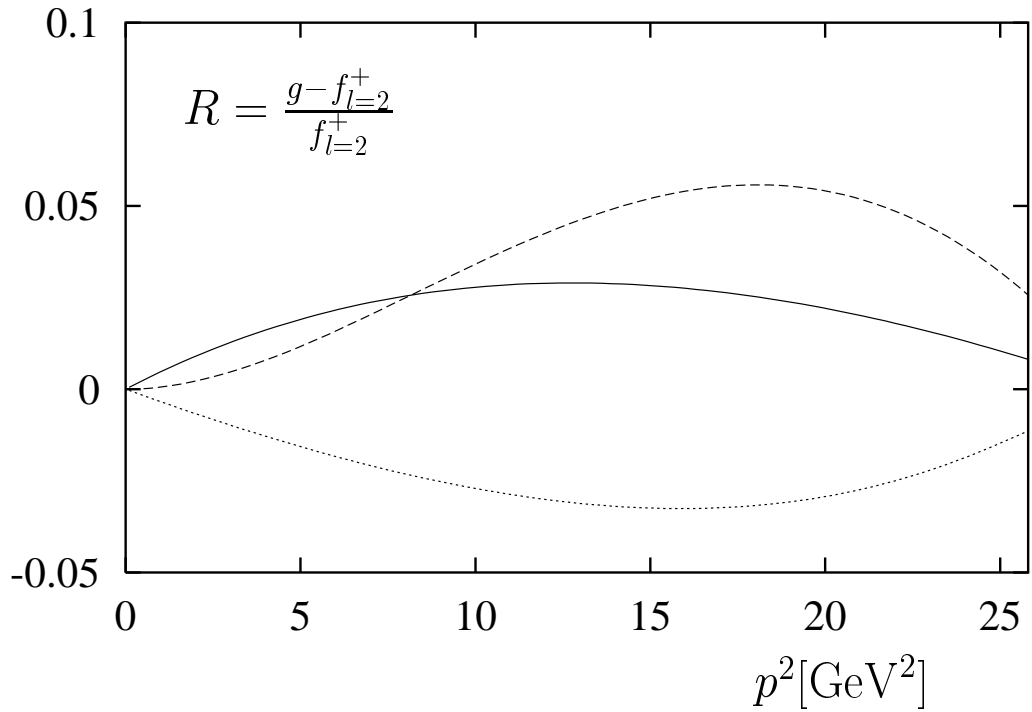,scale=0.9,%
clip=}}
\caption{\label{fig2} 
The relative deviation of the the numerical results with $l=1$ from $l=2$ (solid line),
of $l=3$ from $l=2$ (dashed line) and of the result quoted in \cite{K2000} from
$l=2$ (dotted line).}

\end{figure}

\begin{figure}
\centerline{
\epsfig{file=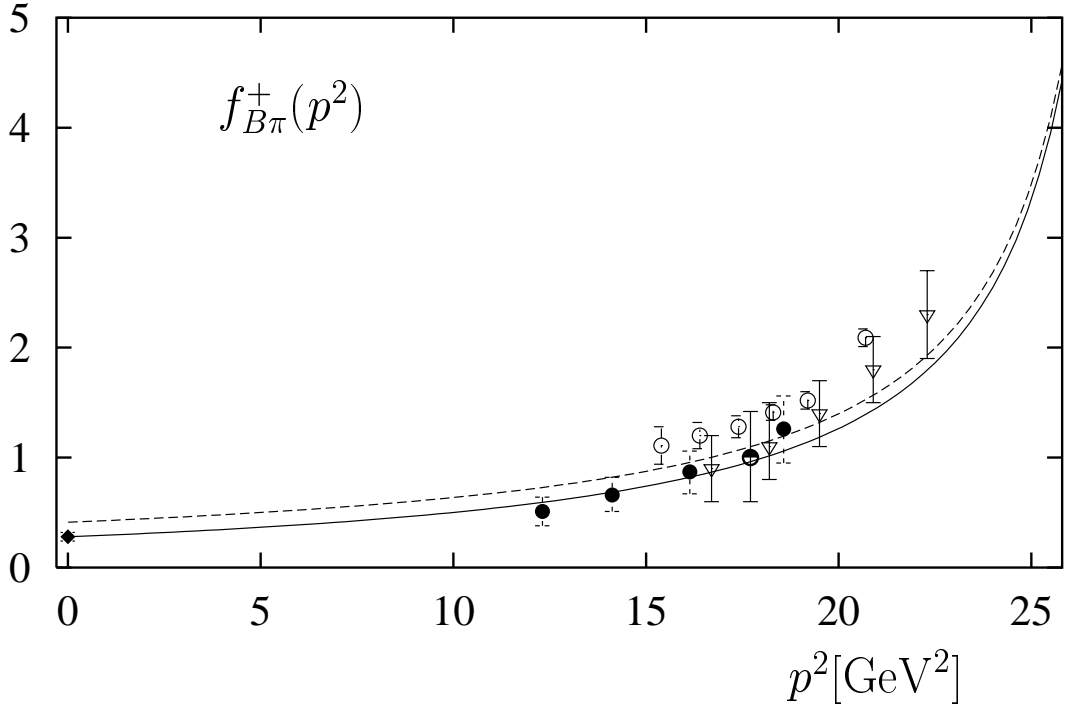,scale=0.9,%
clip=}}
\caption{\label{fig3}
The sum rule results  for the $B\to\pi$ form factor, 
eq.(\ref{eqa6}) with $l=2$ (solid line) and the pole 
model result (dotted line) in comparison to lattice results. 
%The full curves indicate the size of the LCSR uncertainties.
The lattice results come from FNAL \cite{Flynn} (full circles),
UKQCD \cite{UKQCD} (triangles), APE \cite{APE} (full square), 
JLQCD \cite{JLQCD} (open circles), and ELC \cite{Flynn} (semi-full circle).}

\end{figure}

We perform a numerical evaluation of the new sum rule eq.~(\ref{eqa6}) for the values
$l=0,1,2$ and $3$.
We need therefore the values of $f^+$ and its first two derivatives at $p^2=0$.
These numbers can be obtained from the sum rule eq.~(\ref{eqa1a}).
The derivatives are obtained numerically according to
\bq
(f^+)'(0) = \frac{f^+(\Delta s)- f^+(0)}{\Delta s}, & & 
(f^+)''(0) = \frac{f^+(2 \Delta s)-2 f^+(\Delta s) + f^+(0)}{(\Delta s)^2}.
\eq
The sum rules eq.~(\ref{eqa1a}) and eq.~(\ref{eqa2a}) depend on various input parameters,
where each parameter is only known within a certain range.
A complete error analysis, where each parameter was varied within an interval,
was carried out in \cite{K2000}.
In this letter we do not repeat such a complete error analysis.
We only study the dependence on $l$ and we fix here these additional parameters to certain values.
As numerical input parameters we use 
$m_b=4.7\; \mbox{GeV}$, $s_0=35 \;\mbox{GeV}^2$
and  $\alpha(m_Z)=0.118$.
We note that the value of the pole mass is 
compatible with the recently updated $\overline{MS}$ mass
$\bar m( \bar m ) = 4.2 \pm 0.1 $ (for a review of 
recent results see \cite{BenekeHF8}).
The same values of $m_b$ and $s_0$ have been used in the two-point sum 
rule for estimating the decay constant $f_B$.
We obtained $f_B=183 \; \mbox{MeV}$.
The coefficients of the leading twist 
pion distribution amplitude have recently been updated in
\cite{SY,BKM}, the coefficients of the twist 
three and four amplitudes can be found in \cite{BF,BKR,BBKR,balltwist}.
Table~(\ref{t1}) shows the results for the form factor $f^+$ and
its first two derivatives at $p^2=0$ calculated from the 
sum rule eq.~(\ref{eqa1a}).
\begin{table}
\begin{center}
\begin{tabular}{|c|c|c|}
\hline
$f^+_{B\pi}(0)$ & $(f^+_{B\pi})'(0)$ & $(f^+_{B\pi})''(0)$ \\
\hline
$0.28$ & $0.014 \;\mbox{GeV}^{-2}$ & $0.0014 \;\mbox{GeV}^{-4}$ \\
\hline
\end{tabular}
\caption{\label{t1} The form factor $f^+_{B\pi}$ and its first derivatives at $p^2=0\;\mbox{GeV}$.}
\end{center}
\end{table}
We further need 
the $B^*B\pi$ coupling as input data.
This value is obtained from the sum rule eq.~(\ref{eqa2a}) 
as $f_{B^\ast} g_{B^\ast B \pi}=4.4\;\mbox{GeV}$ \cite{BBKR,Coupling,Grozin}. 
With these input values
we evaluate the sum rule eq.~(\ref{eqa6}) for $l=0,1,2$ and $3$. For the Borel parameter
we use $M^2=10 \;\mbox{GeV}^2$.
Our results are shown in fig.~(\ref{fig1}).
First of all we note a remarkable stability of the numerical results with respect to 
changing the number of subtractions $l$.  
The results for $l=1,2$ and $3$ are almost identical.
Secondly, the result for $l=0$ (which is the pole model plus an $\alpha_s$-correction term)
differs at low momentum, but approaches for high momentum the results with subtractions.
This corresponds to the known fact, that for the $B \rightarrow \pi$ transition
the pole model does not describe the form factor accurately at low momentum.
This is also shown in fig.~(\ref{fig3}).
We also note that the standard sum rule for $f^+$ eq.~(\ref{eqa1a}) will differ
significantly from our results at high momentum.
Our results for $l=1,2,3$ agree well
with the parameterization
given in \cite{K2000}:
\be\label{paramB}
f_{B\pi}^+(p^2)=\frac{f_{B\pi}^+(0)}{(1-p^2/m_{B^*}^2)(1-
\alpha_{B\pi}p^2/m_{B^*}^2)} ~, 
\ee
\be
\label{nfplus}
f^+_{B\pi}(0) = 0.28 \pm 0.05, \qquad 
\alpha_{B\pi} =  0.32 \pm ^{0.21}_{0.07}~. 
\ee
To see the difference we take $l=2$ as our main 
result and plot in fig.~(\ref{fig2}) the deviation
\bq \label{diff}
R=\frac{g(p^2)-f^+_{l=2}(p^2)}{f^+_{l=2}(p^2)}
\eq
where we take for $g$ the results $f^+_{l=1}$, $f^+_{l=3}$ 
and the values according to eq.~(\ref{paramB}) with $f_{B\pi}^+(0)=0.28$ and $\alpha_{B\pi} =  0.32$.
The deviations are small. 

In fig.~(\ref{fig3}) we show the sum rule results for the $B\to\pi$ form factor, 
eq.(\ref{eqa6}) with $l=2$ (solid line) 
and the pole model prediction (dotted line) in comparison to lattice results. 
The lattice results come from FNAL \cite{Flynn} (full circles),
UKQCD \cite{UKQCD} (triangles), APE \cite{APE} (full square), 
JLQCD \cite{JLQCD} (open circles), and ELC \cite{Flynn} (semi-full
circle).  
Taking into account the uncertainty of our result, which is estimated to be roughly $15-20\%$ \cite{K2000},
we observe a satisfactory agreement with most lattice results.
However, the results from the JLQCD collaboration lie systematically above our values.
In addition, this group quotes rather small uncertainties.
In general, further improvements in the accuracy of lattice calculations are welcome.

\section{Conclusions} 

In this letter we have shown a method how to obtain the form factor for heavy-to-light
transitions in the whole range of momentum transfer.
Our method extends the QCD sum rule approach and uses
a combination of single and double dispersion relations.
It involves an additional (integer) parameter $l$, corresponding to the number
of subtractions in one channel.
We have derived the corresponding sum rule for the $B\rightarrow \pi$ transition
to twist four accuracy and including radiative corrections to the twist-2 contribution.  
As input data we need the first $(l-1)$ terms of the Taylor expansion of $f^+(p^2)$
around $p^2=0$ as well as the residuum at the pole $p^2=m_{B^\ast}^2$, which can
reliable be obtained from standard sum rules.
The new sum rule involves an additional term, which vanishes in leading order for the
twist two, three and four contributions.
We have calculated the non-vanishing next-to-leading order twist two contribution to this term.
We have shown that variation of the parameter $l$ introduces only small numerical changes in the final
result for the form factor, which are negligible against other uncertainties.\\
\\

{\bf Acknowledgements}\\
\\
We are grateful to M.B. Voloshin, A.I. Vainshtein, R. R\" uckl  
for useful discussions. O.Y. acknowledges 
support from the US Department of Energy (DOE), 
S.W. would like to thank the SPhT Saclay for hospitality.

\newpage


\end{document}